\documentclass[11pt]{article}
\usepackage{vietnam}
\usepackage{amssymb}
\usepackage{url}

\def\Journal#1#2#3#4{{#1} {\bf #2}, #3 (#4)}

\def\be{\begin{equation}}
\def\ee{\end{equation}}
\def\bea{\begin{eqnarray}}
\def\eea{\end{eqnarray}}

\newcommand{\Photo}{\includegraphics[height=35mm]{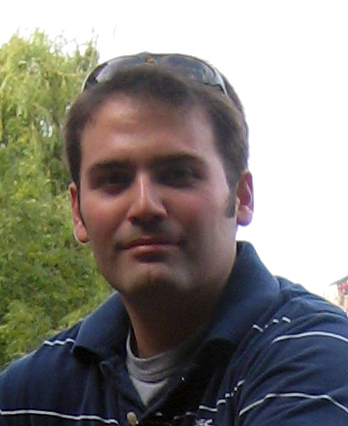}}

\begin{document}
\vspace*{4cm}
\title{The QUIJOTE CMB Experiment: status and first results with the multi-frequency instrument}

\author{M.~L\'opez-Caniego$^{a}$, 
R.~Rebolo$^{b,c,h}$,
M.~Aguiar$^{b}$, 
R.~G\'enova-Santos$^{b,c}$, 
F.~G\'omez-Re\~nasco$^{b}$, 
C.~Gutierrez$^{b}$,
J.M. Herreros$^{b}$, 
R.J.~Hoyland$^{b}$,  
C.~L\'opez-Caraballo$^{b,c}$,  
A.E.~Pelaez Santos$^{b,c}$,
F.~Poidevin$^{b}$, 
J.A. Rubi\~no-Mart\'in$^{b,c}$, 
V.~Sanchez de la Rosa$^{b}$, 
D.~Tramonte$^{b}$,
A.~Vega-Moreno$^{b}$, 
T.~Viera-Curbelo$^{b}$,
R.~Vignaga$^{b}$, 
E.~Mart\'inez-Gonzalez$^{a}$, 
R.B.~Barreiro$^{a}$,
B. Casaponsa~$^{a}$,
F.J.~Casas$^{a}$,
J.M.~Diego$^{a}$, 
R.~Fern\'andez-Cobos$^{a}$, 
D.~Herranz$^{a}$, 
D.~Ortiz$^{a}$,
P.~Vielva$^{a}$, 
E.~Artal$^{d}$, 
B.~Aja$^{d}$, 
J.~Cagigas$^{d}$, 
J.L.~Cano$^{d}$, 
L.~de la Fuente$^{d}$,
A.~Mediavilla$^{d}$,
J.V.~Ter\'an$^{d}$, 
E.~Villa$^{d}$, 
L.~Piccirillo$^{e}$, 
R.~Battye$^{e}$, 
E.~Blackhurst$^{e}$, 
M.~Brown$^{e}$, 
R.D.~Davies$^{e}$,
R.J.~Davis$^{e}$, 
C.~Dickinson$^{e}$,
K.~Grainge$^{f}$,  
S.~Harper$^{e}$, 
B.~Maffei$^{e}$, 
M.~McCulloch$^{e}$, 
S.~Melhuish$^{e}$, 
G.~Pisano$^{e}$, 
R.A.~Watson$^{e}$,
M.~Hobson$^{f}$, 
A.~Lasenby$^{f,g}$, 
R.~Saunders$^{f}$,
and P.~Scott$^{f}$ }

\address{
$^{a}$Instituto de F\'isica de Cantabria, CSIC-UC, Avda. los Castros, s/n, E-39005 Santander, Spain\\
$^{b}$Instituto de Astrof\'isica de Canarias, C/Via Lactea s/n, E-38200 La Laguna, Tenerife, Spain\\
$^{c}$Departamento de Astrof\'{\i}sica, Universidad de La Laguna, E-38206 La Laguna, Tenerife, Spain\\
$^{d}$Departamento de Ingenier\'ia de COMunicaciones (DICOM), Laboratorios de
I+D de Telecomunicaciones, Plaza de la Ciencia s/n, E-39005 Santander, Spain\\
$^{e}$Jodrell Bank Centre for Astrophysics, University of Manchester, Oxford Rd, Manchester M13 9PL, UK\\
$^{f}$Astrophysics Group, Cavendish Laboratory, University of Cambridge, Cambridge CB3 0HE, UK\\
$^{g}$Kavli Institute for Cosmology, University of Cambridge, Madingley Road, Cambridge CB3 0HA\\
$^{h}$Consejo Superior de Investigaciones Cient\'ificas, Spain
}

\maketitle\abstracts{ The {\it QUIJOTE} (Q-U-I JOint Tenerife) {\it CMB} Experiment is designed to 
   observe the polarization of the Cosmic Microwave Background and other 
	Galactic and extragalactic signals at medium and large angular scales in the frequency
	range of 10--40\,GHz. The first of the two QUIJOTE telescopes and 
	the multi-frequency (10--20\,GHz) instrument have been in operation since November 2012. 
	In 2014 a second telescope and a new instrument at 30\,GHz will be 
	ready for commissioning, and  an additional instrument at 40\,GHz is in its final design stages.
	After three years of effective observations, the data obtained by these telescopes and instruments will have the required sensitivity to detect a primordial gravitational-wave 
	component if the tensor-to-scalar ratio is larger than $r=0.05$.  At the moment, we have completed half of the wide Galactic survey with the multi-frequency instrument covering 18\,000 square degrees of the northern hemisphere. When we finish this survey in early 2014, we shall have reached $\sim 14 \mu K$ per one degree beam at 11, 13, 17 and 19 GHz, in both Q and U.}

\def\quijote{{\it QUIJOTE\/}}
\def\Planck{{\it Planck\/}}

\def\eps{\varepsilon}
\def\aap{A\&A}
\def\apj{ApJ}
\def\apjs{ApJS}
\def\apjl{ApJL}
\def\apss{Ap\&SS}
\def\mnras{MNRAS}
\def\aj{AJ}
\def\nat{Nature}
\def\aaps{A\&A Supp.}
\def\prd{Phys. Rev. D}


\section{Introduction}
\label{sec:intro}  
In the last years, experiments observing the Cosmic Microwave Background (CMB) and its anisotropies (e.g., {CBI, Readhead} {\it et al}.~\cite{re}; SPT, {Keisler} {\it et al}.~\cite{ke}; WMAP, Hinshaw {\it et al}.~\cite{hi} and Planck Collaboration 2013 Results. XVI~\cite{pl}) have measured with unprecedented precision many of the parameters that describe our Universe. The CMB and its anisotropies originated $\sim 375.000$ years after the Big Bang, when the Universe had expanded and cooled down sufficiently to allow protons and electrons to form neutral hydrogen atoms. In that epoch, photons started to travel freely until now, when we observe them as a thermal radiation with a black-body spectrum at T $\sim 2.73$ K. Analyses of the CMB suggest that, at a very early time, an exponentially accelerated expansion of the Universe occurred. This inflation period can explain most of the problems that haunted previous cosmological models: the homogeneity and isotropy of the Universe, its nearly flat geometry and the quasi scale invariance of the initial perturbations. Most importantly, theoretical models have allowed scientists to make predictions that can be tested with experiments. A prediction from inflation is that during the accelerated expansion, the quantum fluctuations in the dominant scalar field would have produced both density fluctuations (i.e., scalar fluctuations that have been measured) and a background of gravitational waves (GWB), (i.e., tensor fluctuations). These signals would have left a characteristic imprint on the polarisation of the CMB photons at last scattering that has not been measured yet. 

A very useful tool to synthesize the information contained in the CMB and its anisotropies, both in temperature and polarisation, is the angular power spectrum. This tool quantifies the differences in temperature, from one point of the sky to another, as a function of the angular frequency $\ell$. Theoretical models predict a B-mode component in the angular power spectrum, a combination of primordial B-mode (GWB) and Gravitational Lensing-induced B-mode signal, see Hu \& White~\cite{hu}. The amplitude of the primordial B-mode, given by the ratio of the tensor to scalar fluctuations and quantified with the parameter r, and the parameter $n_{s}$, that characterize the matter density power spectrum, will depend on the model, providing a unique measure of the energy scale of inflation. Moreover, the matter gravitational potential, traced by galaxies, clusters of galaxies and any other large scale structure (LSS) located between us and the last scattering surface, would have also left an imprint on the polarisation of the CMB photons via the gravitational lensing effect (GL-CMB). A detailed knowledge of this lensing induced signal will allow us to probe dark energy parameters, absolute neutrino mass, and even to reconstruct a projected mass map of the LSS. There is an international effort to develop experiments to measure or constrain the amplitude of B-modes power spectrum of the CMB polarisation, and, therefore to attempt to detect primordial gravitational waves. One of these efforts is the \quijote\ experiment, Rubi{\~n}o-Mart\'in {\it et al}.~\cite{ru}, a scientific collaboration between the Instituto de Astrof\'isica de Canarias, the Instituto de F\'isica de Cantabria, the universities of Cantabria, Manchester and Cambridge and the IDOM company,  with the aim of characterizing the polarisation of the CMB, and other galactic and extragalactic emissions at 10--40\,GHz \footnote{http://www.iac.es/proyecto/cmb/pages/en/quijote-cmb-experiment.php}.

\section{Project baseline}
\begin{figure}
\begin{minipage}{0.5\linewidth}
\centerline{\includegraphics[width=1\linewidth]{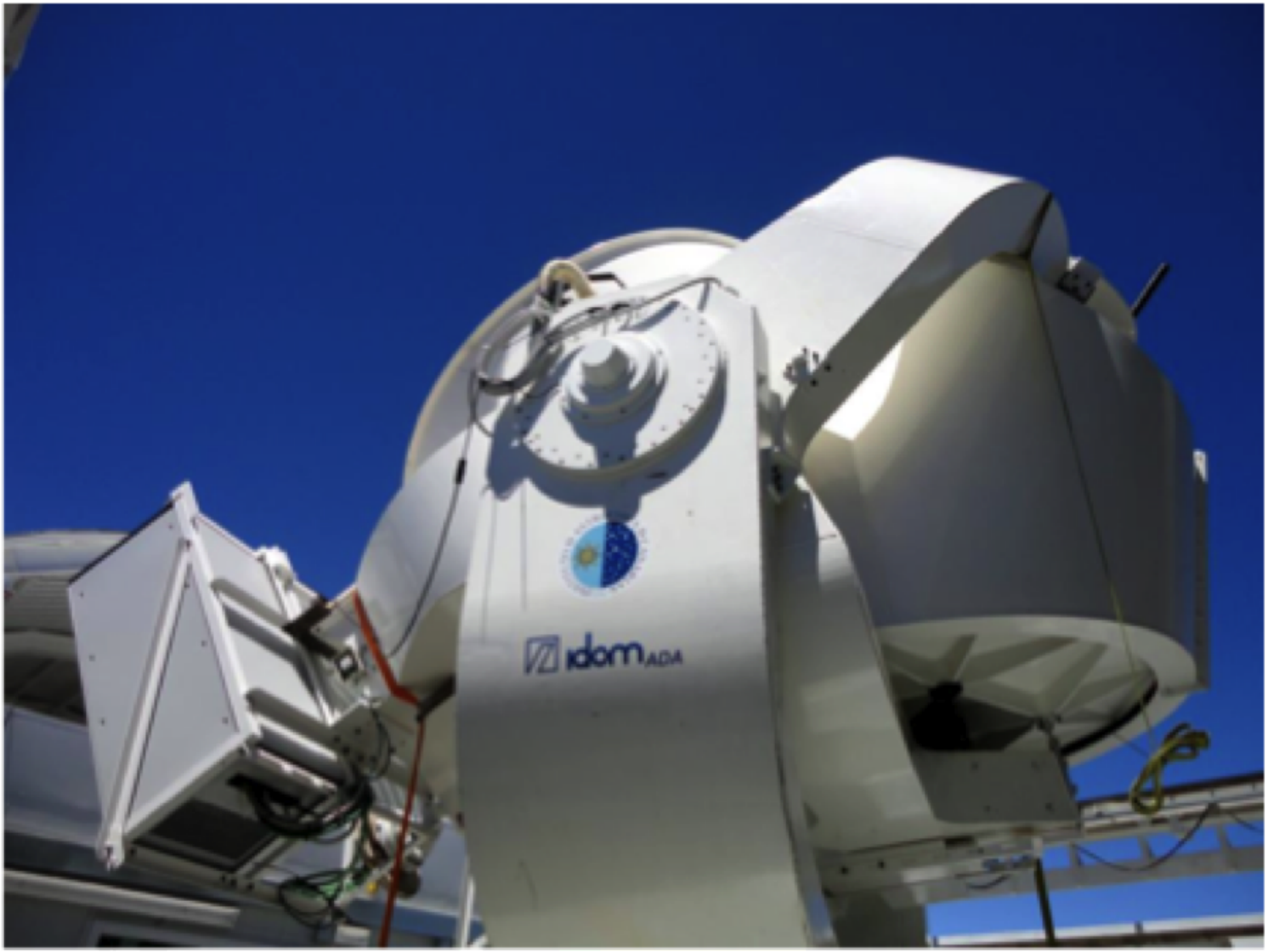}}
\end{minipage}
\begin{minipage}{0.5\linewidth}
\centerline{\includegraphics[width=1\linewidth]{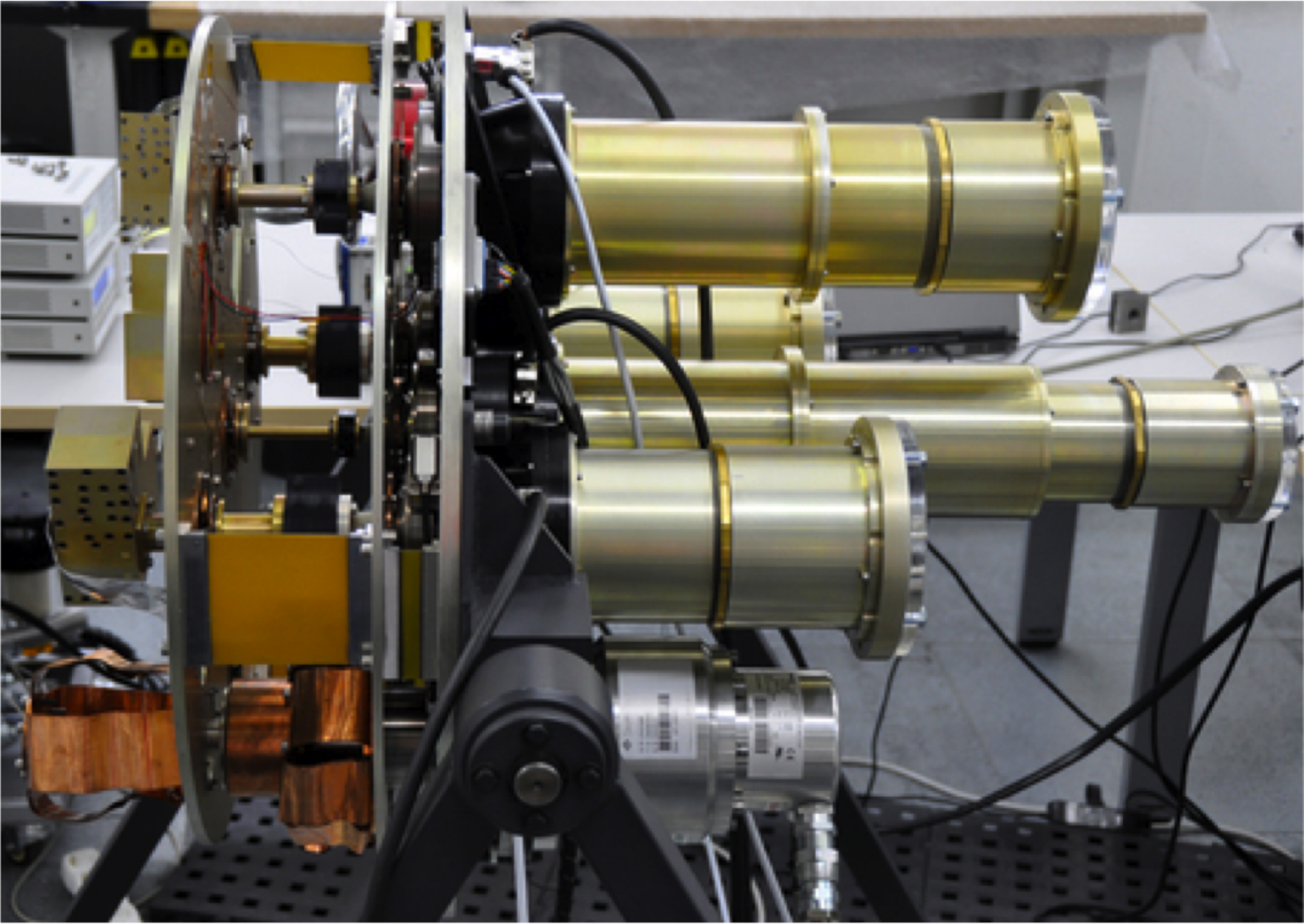}}
\end{minipage}
\caption{Left: The QT1 inside the enclosure at the Teide Observatory. Right: Multi-frequency instrument.}
\label{fig:TELMFI}
\end{figure}

The \quijote\ experiment consists of two telescopes and three instruments which will observe in the frequency range 10--40\,GHz with
an angular resolution of $\sim 1$~degree, from the Teide Observatory (2400\,m)
in Tenerife (Spain). The project has two phases:
\begin{itemize}
\item \textit{Phase I.} A first \quijote\ telescope (QT1) and multi-frequency 
  instrument (MFI), see Figure \ref{fig:TELMFI},  are in operation since November 2012 with a frequency coverage between 10 and 20\,GHz. A second instrument (TGI) with 31 polarimeters working at 30\,GHz is expected to start operations in spring 2014. A 30\,GHz two-element interferometer will monitor and correct the
  contribution of polarized radio-sources in the final \quijote\ maps.
\item \textit{Phase II.} A second \quijote\ telescope (QT2) and a third instrument (FGI) with 40 polarimeters working at 40\,GHz being designed.
\end{itemize}

Table~\ref{tab:basic} summarizes the nominal characteristics of the three instruments in phases I and II. 
\begin{table}
\centering
\caption{Nominal characteristics of the three \quijote\ instruments: MFI, TGI
  and FGI. Sensitivities are referred to Stokes Q and U parameters. See text for details. }
\label{tab:basic}
\begin{tabular}{|l@{ \hspace{0.1cm} }c@{ \hspace{0.1cm} }c@{ \hspace{0.1cm} }c@{ \hspace{0.1cm}
    }c@{ \hspace{0.1cm} }c@{ \hspace{0.8cm} }c@{ \hspace{0.8cm} }|c }
\noalign{\smallskip}\hline
 & \multicolumn{4}{c}{MFI } & TGI & FGI \\ \hline
%
Nominal Frequency [GHz] & 11 &  13 &  17 &  19   & 30 & 40 \\
Bandwidth [GHz]         &  2  &   2 &   2 &   2 & 8 & 10\\
Number of horns         &  2  &   2 &   2 &   2 & 31 & 40 \\
Channels per horn       &  2  &   2 &  2 &   2 &  4 &  4 \\
Beam FWHM [$^\circ$]     & 0.92 &  0.92 &  0.60 &  0.60 & 0.37 & 0.28 \\
$T_{\rm sys}$ [K]         & 25 &  25 &  25 &  25 & 35 & 45\\
NEP [$\mu$K\,$s^{1/2}$] &  395 & 395 & 395 & 395 & 50 & 50 \\
Sensitivity [Jy\,s$^{1/2}$]  & 0.30  &  0.42 &  0.31 &  0.38 & 0.06 & 0.06\\ \hline
\end{tabular}
\end{table}

\section{The Experiment}

\subsection{Telescopes and Enclosure}
The \quijote\ experiment consists of two telescopes installed inside a single enclosure at the Teide Observatory.  The design of
 both QT1 and QT2 telescopes is based on an altazimuth mount
supporting a primary (parabolic) and a secondary (hyperbolic) mirror disposed in
an offset Gregorian Dracon scheme, which provides optimal cross-polarisation
properties ($\le -35$~dB) and symmetric beams. Each primary
mirror has a 2.25\,m projected aperture, while the secondary has 1.89\,m. The
system is under-illuminated to minimize side-lobes, see G\'omez {\it et al}.~\cite{go} . 

\subsection{Instruments}

\subsubsection{Multi-frequency Instrument (MFI)} 
The QUIJOTE MFI is a multi-channel instrument with four independent sky pixels: two operating
at 10--14\,GHz and two at 16--20\,GHz. designed to characterise the Galactic emission. See Hoyland {\it et al}.~\cite{ho} and G\'omez-Re\~nasco {\it et al}.~\cite{go1} for a detailed description. The optical 
arrangement includes four conical corrugated feed horns each of them with a novel cryogenic on-axis
rotating polar modulator which can rotate at speeds of up to 1\,Hz in continuous mode or discrete mode in steps of $22.5^\circ$. All the polarimeters have simultaneous "Q" and "U" detection i.e. the 2 orthogonal linear polar signals are also correlated through
a $180^\circ$ hybrid and passed through two additional detectors. The band
passes of these lower frequency polarimeters have also been split into an upper
and lower band which gives a total of 8 channels per polarimeter (see
Table~\ref{tab:basic}). The receivers use MMIC 6-20\,GHz LNAs, the gain is $\sim$ 30\,dB and the noise temperature $<$ 9\,K.

\subsubsection{Thirty-GHz Instrument (TGI)} 

The QUIJOTE TGI will be devoted to primordial B-mode science.  This instrument will be fitted with 31 polarimeters working at 26-36\,GHz. It includes a fixed polariser and $90^\circ$ and $180^\circ$ phase switches to generate four polarisation states to minimize the different systematics in the receiver. A detailed description of the system has been presented in Cano {\it et al}.~\cite{ca}.

\subsubsection{Forty-GHz Instrument (FGI)} 

The QUIJOTE FGI will be devoted to primordial B-mode science, and will incorporate 40 polarimeters at 40\,GHz. The conceptual design of the polarimeter chain is identical to the TGI.

\subsection{Source Subtractor}

The Cavendish Laboratory and the University of Manchester have upgraded the existing VSA source subtractor (VSA-SS) \cite{VSA1} to monitor the contribution of polarized radio-sources in the \quijote\ maps. The VSA-SS is a two element interferometer, operating at 30\,GHz, with $3.7$\,m dishes and a separation of 9\,m. The upgrade will include a half-wave plate (HWP) in front of each of the antennas in order to allow for successive measurements of Stokes Q and U.  An upgrade of the VSA-SS facility to operate 40\,GHz is under discussion.

\section{Science Program}

\subsection{Core Science}

The \quijote\ experiment has two primary scientific goals:
\begin{itemize}
\item to detect the imprint of gravitational B-modes if they have an amplitude
  $r \ge 0.05$.
\item to provide essential information of the polarisation of the synchrotron
  and the anomalous microwave emissions from our Galaxy at low frequencies
  (10--40\,GHz).
\end{itemize}
For this purpose, \quijote\ is conducting two large surveys in
polarisation (i.e. Q and U):

\hspace*{0.3cm} i) \textbf{Cosmological Survey}: deep survey covering $\sim$ 3\,000\,deg$^2$ of the sky.  Here, we shall reach sensitivities of $\sim 5$\,$\mu$K 
per one degree beam after two years of effective observing time with the MFI (11--19\,GHz), and $\lesssim 1$\,$\mu$K per beam with TGI and FGI at 30 and 40\,GHz. At the moment we have accumulated $\sim$ 1200 hours of observing time in these fields.

\begin{figure*}
\begin{minipage}{0.5\linewidth}
\centerline{\includegraphics[width=1\linewidth]{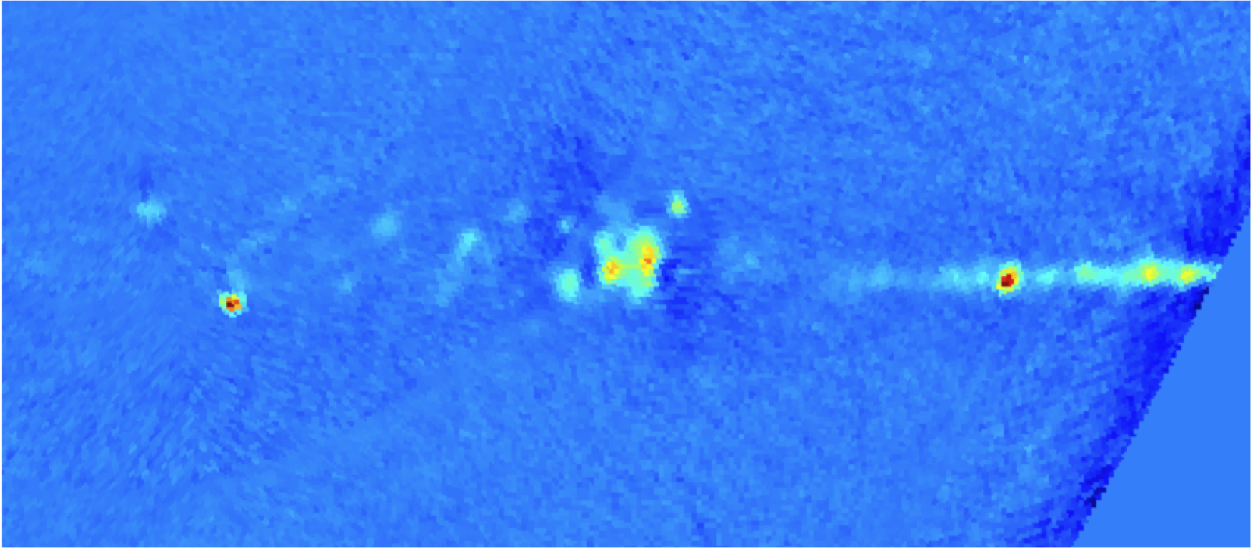}}
\end{minipage}
\begin{minipage}{0.5\linewidth}
\centerline{\includegraphics[width=1\linewidth]{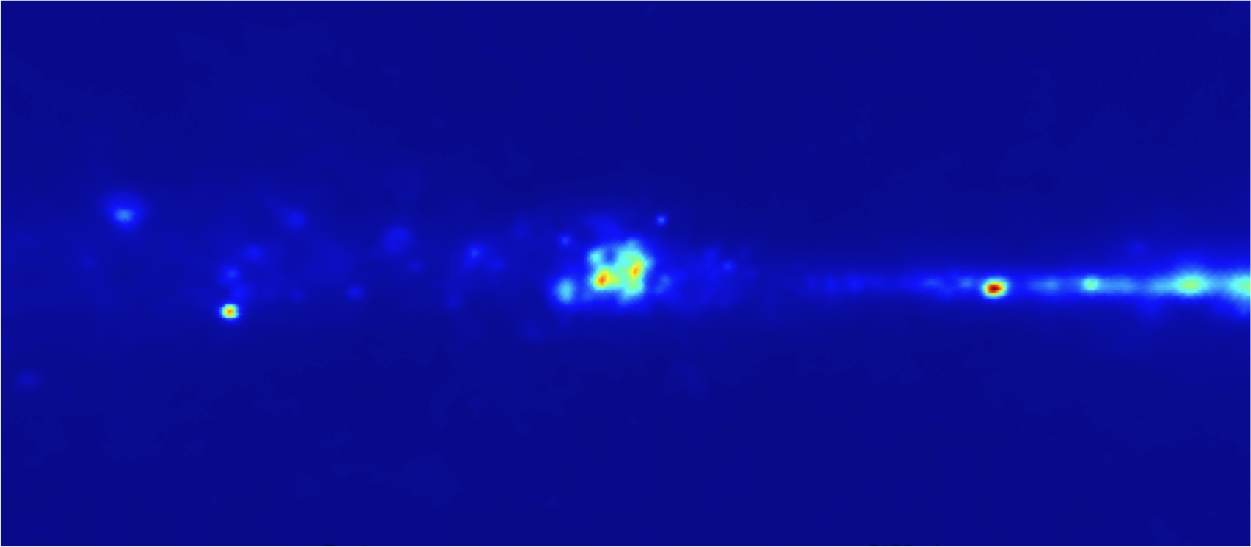}}
\end{minipage}
\caption{Left: close-up view of the \quijote\ 11 GHz wide Galactic survey intensity map centred on the Cygnus X region from ten days of data of the survey. Right: WMAP nine year coadded intensity map at 23 GHz covering the same region of the sky. The two bright compact regions to the left and to the right of the Cygnus X are the supernova remnants CasA and W51, respectively.}
\label{fig:cygnus}
\end{figure*}

\hspace*{0.3cm} ii) \textbf{Wide Galactic Survey}: shallow survey covering 18\,000\,deg$^2$ of sky.  Half of the survey was completed between April and July with the MFI (11--19\,GHz), see Fig.  \ref{fig:cygnus}. In the next months we will repeat the observations increasing the integration time a factor 2, reaching sensitivities of $\sim
$14\,$\mu$K per one degree beam in the Stokes Q and U maps. In the future, a similar survey will be conducted with the TGI and FGI, expecting to reach sensitivities of $\lesssim 3$\,$\mu$K per beam.

According to these nominal sensitivities, \quijote\ will provide one of the most
sensitive 11--19\,GHz measurements of the polarisation of the synchrotron and
anomalous emissions on degree angular scales. This information is extremely
important given that B-modes are known to be inferior in amplitude when 
compared to the Galactic emission, see Tucci {\it et al}.~\cite{tu}. The \quijote\ maps will also constitute a unique
complement of the \Planck\ \,satellite~\footnote{\Planck:
  \url{http://www.rssd.esa.int/index.php?project=Planck}}, helping in the
characterisation of the Galactic emission. 

In particular, the combination of \Planck\ and \quijote\ will allow us to determine synchrotron spectral
indices with high accuracy, and to fit for curvature of the synchrotron spectrum
to constrain CR electron physics, Kogut {\it et al}.~\cite{ko}; to study the large-scale
properties of the Galactic magnetic field, Ruiz-Granados {\it et al}.~\cite{ru2} and to assess the level
of a contribution of polarized anomalous microwave emission, Watson {\it et al}.~\cite{wa} and Battistelli {\it et al}.~\cite{ba}.

\begin{figure}
\begin{minipage}{1\linewidth}
\centerline{\includegraphics[width=1\linewidth]{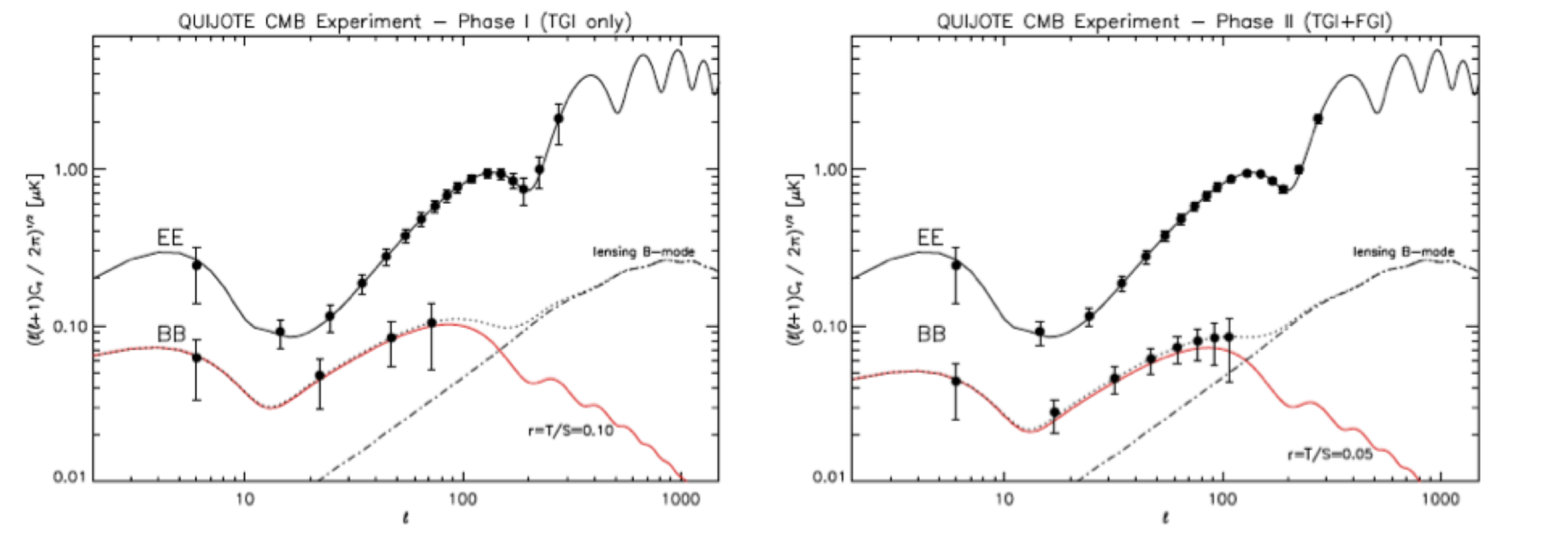}}
\end{minipage}
\caption{Left: Example of the \quijote\ scientific goal after the Phase I of the
  project, for the angular power spectrum of the CMB E and B mode signals. It is
  shown the case for 1~year (effective) observing time, and a sky coverage of
  $\sim 3,000$\,deg$^2$. The red line corresponds to the primordial B-mode
  contribution in the case of $r=0.1$. Dots with error bars correspond to
  averaged measurements over a certain multipole band. Right: Same computation
  but now for the \quijote\ Phase II. Here we consider 3 years of effective
  operations with the TGI, and that during the last 2 years, the FGI will be
  also operative. The red line now corresponds to $r=0.05$. This figure was originally published 
  in Rubi\~no-Mart\'in {\it et al}.~\protect\cite{ru2}.}
\label{fig:goal}
\end{figure}

Using the MFI maps from the deep survey, we plan to correct the high frequency
\quijote\ channels (30 and 40\,GHz) to search for primordial B-modes. As an
illustration, the left panel of Fig.~\ref{fig:goal} shows the scientific goal for the angular power spectrum of the E and B modes after 1-year
of effective observing time, assuming a sky coverage of 3\,000 square degrees,
with the TGI. In this case, the final noise level for the 30\,GHz map is $\sim 0.5$\,$\mu$K/beam.  The right panel shows the scientific
goal for the \quijote\ Phase II. Here, we consider 3 years of effective
observing time with the TGI, and 2 years with the FGI. Note that the TGI and FGI can operate simultaneously, one in each telescope.  Finally, we stress that the computations presented
in Fig.~\ref{fig:goal} correspond to the optimal situation in which the
foreground removal leaves a negligible impact on the power spectrum. 

\subsection{Non-Core Science}

In addition to the core-science goals, the characteristics of
\quijote\ make it a suitable experiment for performing (relatively-short)
observations in specific regions of the sky that would allow us to tackle other scientific
objectives such as:

\hspace*{0.3cm} i)\hspace*{0.1cm} Study of the polarisation of Galactic regions
and extragalactic sources. 

\hspace*{0.3cm} ii)\hspace*{0.1cm} Study of the North Polar Spur. This a huge
feature, visible mainly in radio wavelengths, which covers about a quarter of
the sky and extends to high Galactic latitudes. 

\hspace*{0.3cm} iii)\hspace*{0.1cm} Study of the polarisation of the anomalous
microwave emission (AME) in the Perseus molecular cloud, already observed for 150 hours covering $\sim 350$\,deg$^2$, and other 
Galactic clouds. 

\hspace*{0.3cm} iv) Study of the Galactic haze in polarisation, an excess signal towards the Galactic centre seen in WMAP and Planck, with a significantly flatter spectrum than synchrotron and gamma-ray counterpart in Fermi, Dobler {\it et al}.~\cite{do}, already observed for 160 hours covering $\sim 1200$\,deg$^2$.

\hspace*{0.3cm} v) Study of the polarisation of the cold spot. This is a
non-Gaussian feature in the CMB, in the form of an extremely extended and cold
region, which was found in WMAP and Planck data, Vielva {\it et al}.~\cite{vi} and Planck Collaboration Results XXIII \cite{plxxiii}

\section{Project status}

QT1 and MFI have been operating since November 2012, and we expect to commission the QT2 and TGI in the spring of 2014. The FGI will be ready for commissioning in early 2015.

\section*{Acknowledgements} 
The \quijote\ experiment is being developed by the Instituto de Astrofisica de
Canarias (IAC), the Instituto de F\'isica de Cantabria (IFCA), and the
Universities of Cantabria, Manchester and Cambridge.  Partial financial support
is provided by the Spanish Ministry of Economy and Competitiveness (MINECO)
under the projects AYA2012-39475-C01 and C02 and by the Consolider-Ingenio 
project CSD2010-00064 (EPI: Exploring the Physics of Inflation).


\section*{References}

\begin{thebibliography}{99}

\bibitem{re} A.~C.~S. Readhead  {\it et al}, \Journal{\apj}{609}{498}{2004}.
\bibitem{ke} R. Keisler {\it et al}, \Journal{\apj}{743}{28}{2011}.
\bibitem{hi} G. Hinshaw {\it et al}, \Journal{arXiv1212.5226}{}{}{2012}.
\bibitem{pl} Planck Collaboration Results 2013 XVI \Journal{arXiv1303.5076P}{}{}{2013}.
\bibitem{hu} W. Hu and M. White , \Journal{New Astronomy}{2}{323}{1997}.
\bibitem{ru} J.~A Rubi\~no-Mart\'in {\it et al}, \Journal{Highlights of Spanish Astrophysics}{V}{}{2010}.
\bibitem{go} A. G\'omez  {\it et al}, \Journal{SPIE Conference Series}{7733}{29}{2012}.
\bibitem{ho} R. Hoyland  {\it et al}, \Journal{SPIE Conference Series}{8452}{33}{2012}.
\bibitem{go1} F. G\'omez-Re\~nasco  {\it et al}, \Journal{SPIE Conference Series}{8452}{34}{2012}.
\bibitem{ca} J. L. Cano {\it et al}, \Journal{Proc. of the 42nd European Microwave Conference}{}{37}{2012}.
\bibitem{VSA1} R.~A. Watson {\it et al}, \Journal{\mnras}{341}{1057}{2003}.
\bibitem{tu} M. Tucci {\it et al}, \Journal{\mnras}{360}{935}{2005}.
\bibitem{ko} A. Kogut, \Journal{\apj}{753}{110}{2012}.
\bibitem{ru2} B. Ruiz-Granados  {\it et al}, \Journal{\aap}{522}{73}{2010}.
\bibitem{wa} R.~A. Watson {\it et al}, \Journal{\apjl}{624}{89}{2005}.
\bibitem{ba} E.~S. Battistelli {\it et al}, \Journal{\apjl}{645}{141}{2006}.
\bibitem{ru} J.~A Rubi\~no-Mart\'in {\it et al}, \Journal{SPIE Conference Series}{8444}{}{2012}.
\bibitem{do} G. Dobler {\it et al}, \Journal{\apj}{750}{17}{2012}.
\bibitem{vi} P. Vielva {\it et al}, \Journal{\apj}{609}{22}{2004}.
\bibitem{plxxiii} Planck Collaboration Results 2013 XXIII \Journal{arXiv1303.5083}{}{}{2013}.

\end{thebibliography}
\end{document}